\documentclass{WileyMSP-template}
\usepackage{xcolor}

\begin{document}

\pagestyle{fancy}


\title{Efficient Frequency Doubling with Active Stabilization on Chip} 

\maketitle


\author{Jia-Yang Chen*}
\author{Chao Tang}
\author{Mingwei Jin}
\author{Zhan Li}
\author{Zhaohui Ma}
\author{Heng Fan}
\author{Santosh Kumar}\\
\author{Yong Meng Sua}
\author{Yu-Ping Huang*}



\begin{affiliations}
Dr. Jia-Yang Chen, Chao Tang, Mingwei Jin, Zhan Li, Zhaohui Ma, Heng Fan, Dr. Santosh Kumar, Dr. Yong Meng Sua, Prof. Yu-Ping Huang\\
Address:Department of Physics, Stevens Institute of Technology, 1 Castle Point Terrace, Hoboken, New Jersey, 07030, USA\\
Center for Quantum Science and Engineering, Stevens Institute of Technology, 1 Castle Point Terrace, Hoboken, New Jersey, 07030, USA\\
Email: jchen59@stevens.edu\\
Corresponding Email: yhuang5@stevens.edu

\end{affiliations}


\begin{abstract}
Thin-film lithium niobate (TFLN) is superior for integrated nanophotonics due to its outstanding properties in nearly all aspects: strong second-order nonlinearity, fast and efficient electro-optic effects, wide transparency window, and little two photon absorption and free carrier scattering. Together, they permit highly integrated nanophotonic circuits capable of complex photonic processing by incorporating disparate elements on the same chip. Yet, there has to be a demonstration that synergizes those superior properties for system advantage. Here we demonstrate such a chip that capitalizes on TFLN's favorable ferroelectricity, high second-order nonlinearity, and strong electro-optic effects. It consists of a monolithic circuit integrating a Z-cut, quasi-phase matched microring with high quality factor and a phase modulator used in active feedback control. By Pound–Drever–Hall locking, it realizes stable frequency doubling at about 50\% conversion with only milliwatt pump, marking the highest by far among all nanophotonic platforms with milliwatt pumping. Our demonstration addresses a long-outstanding challenge facing cavity-based optical processing, including frequency conversion, frequency comb generation, and all-optical switching, whose stable performance is hindered by photorefractive or thermal effects. Our results further establish TFLN as an excellent material capable of optical multitasking, as desirable to build multi-functional chip devices. 
\end{abstract}

\section{Introduction}
The efficiency of photonic integrated circuits can be elevated by orders of magnitude through cavity enhancement, giving rise to ultra-efficient frequency doubling \cite{chen2019ultra, lu2019periodically, lu2020toward},  optical parametric oscillation \cite{lu2019milliwatt, bruch2019chip, lu2021ultralow}, frequency comb generation \cite{wang2019monolithic, he2019self, bruch2021pockels}, quantum photon sources \cite{lu2019chip, ma2020ultrabright}, and quantum frequency conversion \cite{singh2019quantum}, etc. They promise future chip devices capable of classical and quantum optical processing at large scale and with high modularity. However, any cavity enhancement necessarily comes at the price of aggravated photorefractive (PR) and thermal effects, which challenges, if not preventing, achieving the device's due performance over time. For example, thin-film lithium niobate (TFLN) microrings can achieve ultrahigh normalized efficiency for second-harmonic generation (SHG), with $\eta_\mathrm{nor} = P_\mathrm{sh}/P^2_\mathrm{p}$ ($P_\mathrm{p(sh)}$ being the pump (second-harmonic) power on chip) exceeding 100$\%$ mW$^{-1}$ \cite{chen2019ultra, lu2019periodically, lu2020toward, ma2020ultrabright}. However, the absolute conversion efficiency $\eta_\mathrm{abs} = P_\mathrm{sh}/P_\mathrm{p}$, which is a critical metric benchmarking the end system performance, is capped at $20\%$ due to thermal, PR, and Zeno blockade effects \cite{Huang:13, chen2017observation}. This deficiency must be overcome before cavity-enhanced TFLN devices can be deployed for practical applications. 

Here, we demonstrate a scalable and overhead-friendly approach to overcoming the detrimental PR and thermal effects in a high quality factor cavity. Aiming specifically at testing the prospect of creating highly integrated and self-stabilized TFLN devices, we incorporate the cavity with a phase modulator on the same chip. Thanks to TFLN's exceptional ferroelectric, nonlinear optical, and electro-optic properties, we achieve over 51 $\%$ absolute SHG efficiency with only 1.8 mW pump power, scoring a new high among all nanophotonic platforms at such milliwatt level driving power \cite{chen2019ultra, lu2019periodically, bruch2019chip, ma2020ultrabright, lu2019efficient, logan2018400, chang2019strong}. Assisted by the on-chip phase modulator, we demonstrate stable SHG with conversion efficiency around $48\%$ for over 30 minutes by Pound–Drever–Hall (PDH) locking. This is an essential step toward practical nonlinear and quantum photonics applications based on lithium niobate microring devices.


\begin{figure}
  \includegraphics[width=\linewidth]{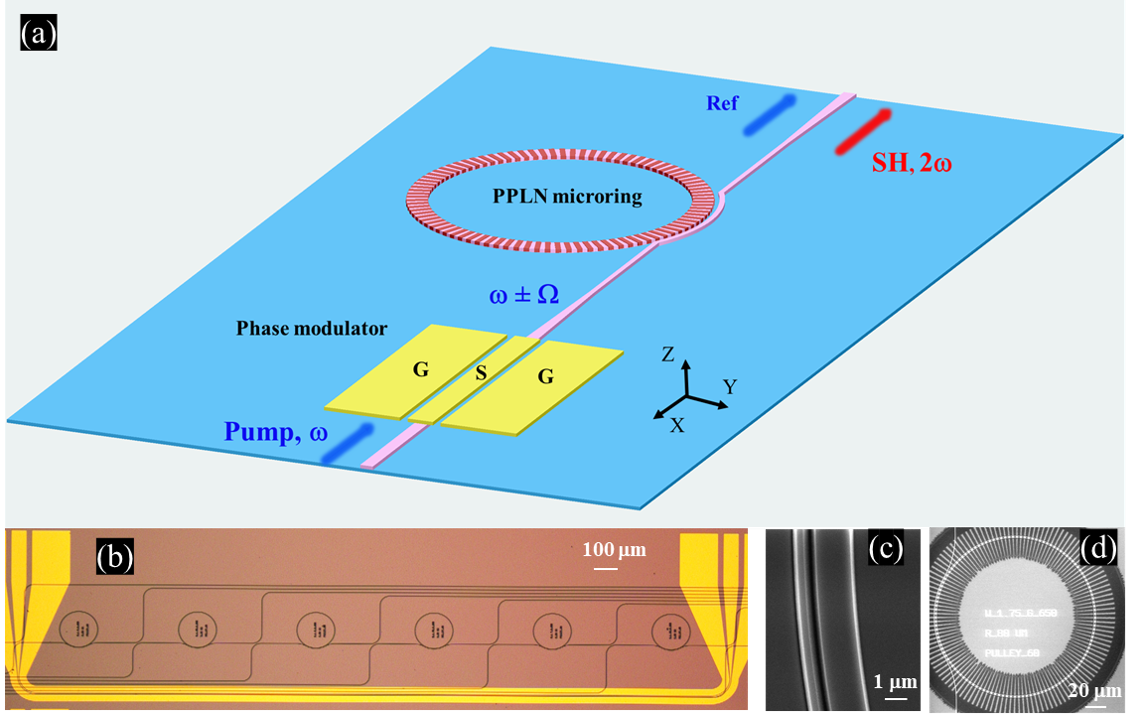}
  \caption{Integrated TFLN circuits for frequency doubling and stabilization. (a) Schematic consisting of an electro-optic phase modulator and a Z-cut, periodically poled microring. High-speed modulation signal ($\Omega$) is applied on the electrodes to generate sidebands around the pump's carrier frequency ($\omega$ $\pm$ $\Omega$). The modulated pump then passes through the microring to generate error signal with an off-chip detector and electronic circuits for feedback control. A pulley coupler is designed for efficient coupling both infrared and visible band. In the same microring, the pump could also efficiently generate second-harmonic light at $2\omega$ via $\chi^{(2)}$ parametric conversion. (b) shows the microscope image for the electrodes on top of the fabricated structures. (c) and (d) are the SEM images of the etched pulley coupler before poling process and the poled microring structure after removing the poling electrodes, respectively.}
  \label{fig0}
\end{figure}

\section{Device design and fabrication}
To maximize the absolute conversion efficiency ($\eta_\mathrm{abs}$), one will need to improve the extraction efficiency (i.e., the microring to bus waveguide coupling) for the periodic poled lithium niobate (PPLN) microring. Assuming quasi-phase matching and no cavity detuning, the highest absolute conversion efficiency is given by \cite{lu2019periodically, breunig2016three}:
\begin{equation}
  \eta_\mathrm{abs, max} = \frac{Q_\mathrm{p,L}}{Q_\mathrm{p,c}} \frac{Q_\mathrm{sh,L}}{Q_\mathrm{sh,c}},
  \label{eq1}
\end{equation}
where $Q_{m,n}$ is the quality factor, with $m$=p, sh standing for the pump and second-harmonic cavity modes, respectively, and $n$=c,L denoting the coupling and loaded Q. The maximum conversion efficiency can be reached at pump power of \cite{lu2019periodically, breunig2016three}:
\begin{equation}
P_p = 16~ \frac{\eta_\mathrm{abs, max}}{\eta_\mathrm{nor}}
\end{equation}
Thus, high normalized efficiency $\eta_\mathrm{nor}$ (e.g., $\sim$150$\%$ mW$^{-1}$) is important to achieve high absolute efficiency $\eta_\mathrm{abs}$ (e.g., $\sim$50$\%$) with only milliwatt pump power.

For quasi phase matching, concentric periodic-poling is applied to the microring resonator to attain the strongest possible interaction between the fundamental quasi-transverse-magnetic (quasi-TM) cavity modes in both infrared (IR) and visible bands, by realizing ideal mode overlapping and utilizing the largest nonlinear tensor $d_{33}$ in lithium niobate. Meanwhile, we optimize the dimensions of the pulley coupler to satisfy over-coupling condition for both IR and visible cavity modes, as Eq.(\ref{eq1}) requires. To this end, the top-width ($w_{pulley}$) of the pulley coupler is first determined to satisfy the phase matching condition for quasi-TM$_{00}$ modes in visible band, with $n_{pulley} R_{pulley} = n_{ring} R_{ring} $ \cite{hosseini2010systematic}, where $n_{pulley}$ and $n_{ring}$ denote the effective refractive indices of the waveguide and cavity modes, respectively, and $R_{pulley}$ and $R_{ring}$ denote the radii of the pulley coupler and microring. For the visible cavity mode, the coupling strength $\kappa_{c,vis}$ is increased when the length of the pulley coupler, $L_{pulley}$ is longer and the gap  between the pulley coupler and microring, $g_{pulley}$, is smaller. For the IR cavity mode, however, the coupling strength $\kappa_{c, IR}$ oscillates as $\mathrm{sinc}(\Delta\Phi)$, when the phase mismatch $\Delta\Phi$ is large. Because of the two distinct behaviors, the length $L_{pulley}$ and the gap $g_{pulley}$ can be varied to optimize the coupling strength for both bands. Finally, to stabilize SHG in high pump power region, we integrate an on-chip phase modulator driven by a gigahertz radio wave, by taking advantage of TFLN's largest electro-optic tensor $r_{33}$ to create error signal for feedback control, as detailed later in Section \ref{PDH}. \\

The schematic of our device is shown in Fig.~\ref{fig0}. The entire device is fabricated on a Z-cut LNOI wafer (NANOLN Inc.), with a 700-nm thick LN thin film bonded on 2 $\mu$m silicon dioxide layer above a silicon substrate. First, the microring and waveguide structure are defined by electron beam lithography (EBL). The top width and the radius of the microring are 1.75 $\mu$m and 80 $\mu$m, respectively. A slightly larger radius is chosen here in consideration of potential bending loss due to shallower etching and worse sidewall angle compared to 55 $\mu$m in \cite{ma2020ultrabright}. Then, ion milling is applied to shallowly etch the structures, where 410-nm thick LN is etched with 290-nm LN remaining, and the sidewall angle is approximately 67$^\circ$. The optimized pulley coupler, shown in Fig.~\ref{fig0}(c) with $w_{pulley}$ = 400 nm, $g_{pulley}$ = 650 nm, and $L_{pulley}$ = 80 $\mu$m, is created to increase the ring-bus waveguide coupling to attain simultaneously over-coupling condition for both the visible and IR lightwaves. Then, a concentric periodically poled region with a period of 4.0 $\mu$m (see Fig.~\ref{fig0}(d)) is created via several 1-ms and 525-V electrical pulses using a similar process described in \cite{chen2020efficient}. After removing the poling electrodes and cladding with 1 $\mu$m silicon dioxide, we carry out the following EBL process to define the phase modulator structure and deposit 30-nm Cr and 300-nm Au as the electrode material via electron beam evaporator. The phase modulator has a standard ground-signal-ground (GSG) electrode structure, as shown in Fig.~\ref{fig0}(b). The top width of the waveguide of the phase modulator is 1.5 $\mu$m. The width of the signal pad and ground pad are 5 $\mu$m and 50 $\mu$m, respectively, with 3 $\mu$m gap between them. The width of the signal pad is chosen under the trade-off between impedance matching and modulation efficiency, where a wider signal pad width ($>$ 10 $\mu$m) helps achieving 50 $\ohm$ impedance but reduces effective overlap between electrical field with optical mode in the LN waveguide. In future work, we can further optimize the dimension of the signal pad to achieve both efficient modulation and 50 $\ohm$ impedance. The effective modulation length of the electrode is 0.14 cm. In the following section, we will first individually characterize each component and then study the overall performance through the feedback experiment. 

\section{Experimental results}
\begin{figure}
  \includegraphics[width=\linewidth]{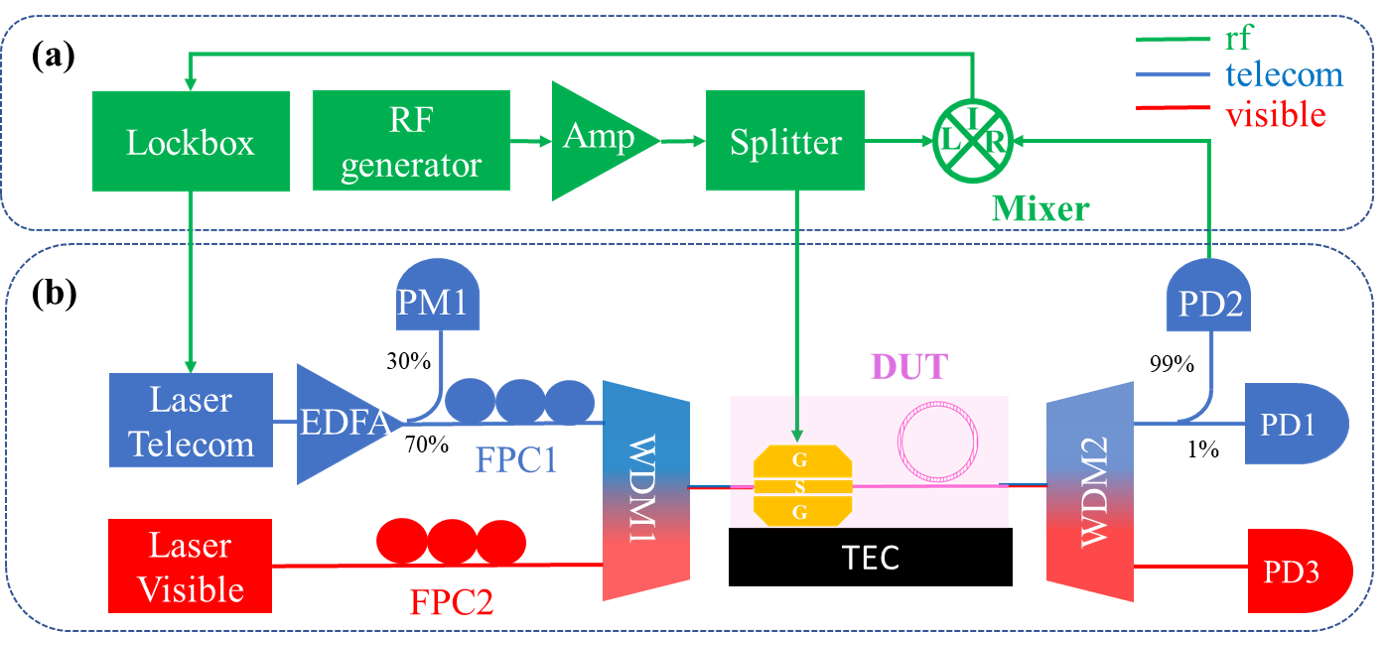}
  \caption{Experimental schamtics, with (a) the Pound–Drever–Hall locking configuration and (b) linear and nonlinear optical setup. The green, blue and red lines denote the radio wave channel, telecome light path and visible path, respectively. FPC: Fiber Polarization Controller; EDFA: erbium doped fiber amplifier; TEC: temperature electronic controller;  WDM 1(2): telecom-visible wavelength-division multiplexing module; RF: radio-frequency; PM: power meter; PD: photodetector; DUT: device under test; Amp: electrical amplifier.}
  \label{fig1}
\end{figure}
\subsection{Second-harmonic generation in a PPLN microring resonator}
\begin{figure}
  \includegraphics[width=\linewidth]{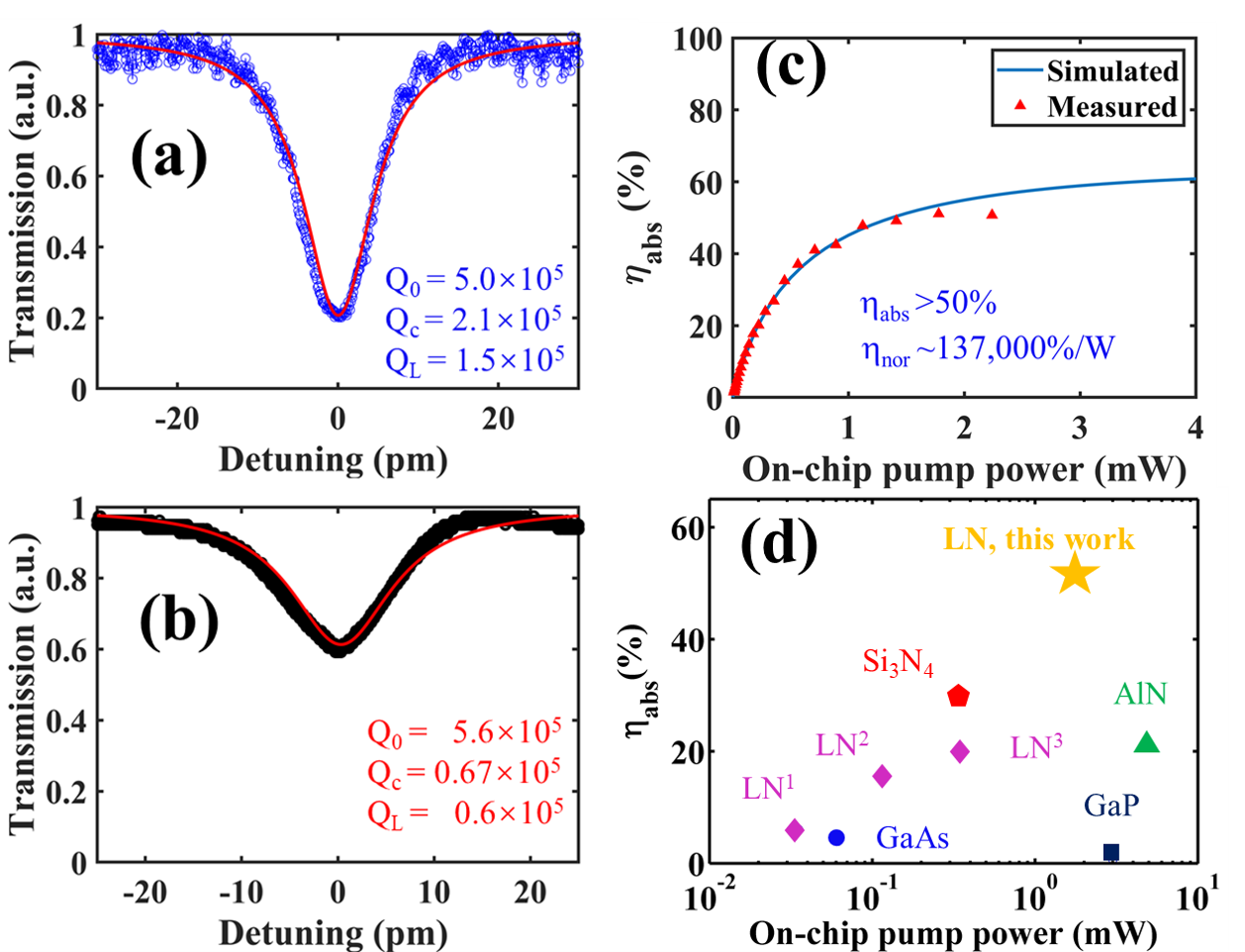}
  \caption{ (a) and (b) Spectra of IR and visible TM$_{00}$ cavity modes around 1561.4 nm and 780.7 nm, respectively. (c) Absolute SHG efficiency $\eta_\mathrm{abs}$ versus the on-chip pump power. (d) Absolute SHG efficiency and the corresponding required pump power demonstrated in various integrated $\chi^{(2)}$ and $\chi^{(3)}$ photonics platforms, including  LN$^1$ \cite{chen2019ultra}, LN$^2$ \cite{lu2019periodically}, LN$^3$\cite{ma2020ultrabright}, AlN \cite{bruch2019chip}, Si$_3$N$_4$ \cite{lu2019efficient}, GaP \cite{logan2018400} and GaAs \cite{chang2019strong}.
 In (d) we have only included results with milliwatt or sub-millwatt pump power, in considering future power-efficient applications. There are demonstrations of 60$\%$ at 60 mW pump power in a Si$_3$N$_4$ microring  \cite{li2016efficient} and 42$\%$ at 35 mW pump power in a AlN microring \cite{wang2020efficient}. Their absolution conversion will be rather low at milliwatt pump power.}
  \label{fig2}
\end{figure}

We first characterize the linear optical properties of the device. As shown in Fig.~\ref{fig1} (b), two polarized tunable lasers (Santec 550, 1500-1630 nm and Newport TLB-6712, 765-781 nm) and tapered fibers (OZ OPTICS) are used to independently characterize fiber-chip-fiber coupling, whose losses are measured to be 17 dB around 1560 nm and 16 dB around 780 nm, respectively. Using the same setup, we obtain the loaded ($Q_\mathrm{L}$), coupling ($Q_\mathrm{c}$) and intrinsic ($Q_\mathrm{0}$) quality factors for both bands, where $Q_\mathrm{L} (Q_\mathrm{c},Q_\mathrm{0}) \approx 1.5 (2.1, 5.0)\times 10^5$ for the IR mode and $Q_\mathrm{L} (Q_\mathrm{0}) \approx 0.6 (0.67, 5.6)\times 10^5$ for the visible mode, as shown in Fig.~\ref{fig2} (a) and (b). According to Eq.~(\ref{eq1}), over-coupled condition at both wavelengths are essential to achieve high absolute conversion efficiency. For the current device, the highest possible conversion efficiency is $\eta_\mathrm{abs, max}=64\%$. For near unity efficiency, the cavity needs to be further overcoupled for both the visible and IR modes. \\

We then characterize the device's nonlinearity using second-harmonic generation. By sweeping the infrared pump laser and fine tuning the device's temperature, a quasi-phase matched SHG is achieved for optimal resonance modes at 1561.4 nm and its second-harmonic at 780.7 nm around 75.4 $^\circ C$. The temperature dependency of the SHG efficiency shows an error tolerance of 1.1$^\circ C$. To extracted the normalized and absolute conversion efficiency, we gradually increase the on-chip pump power from microwatt to millwatt level, as shown in Fig.~\ref{fig2} (c). In the undepleted region, with 79.4 $\mu$W pump power in the input fiber, 27.5 nW SH power is collected in the output fiber. Accounting for the coupling loss and considering negligible propagation loss, the normalized SHG efficiency $\eta_\mathrm{nor}$ is estimated to be $137,000 \%/W$. In the depleted region, with 12.6 mW pump power in the input fiber, 144 $\mu$W SH power is collected in the output fiber. The absolute SHG efficiency $\eta_\mathrm{nor}$ saturates at 51$\%$ with 1.8 mW on-chip pump power after correcting for the coupling loss. Figure~\ref{fig2} (d) provides a comparison of the absolute conversion efficiency and its required pump power across the state-of-the-art $\chi ^{(2)}$ and $\chi ^{(3)}$ integrated photonic platforms. With high absolute conversion efficiency ($\sim50\%$) achieved at low pump power ($\sim$mW), our device is outstanding for quantum frequency conversion and beyond, promising  ultrahigh efficiency and noise as desirable for system integration \cite{singh2019quantum,fan2021photon}.\\

Although the conversion efficiency for the current device can theoretically reach 64$\%$, in practice the attainable $\eta_\mathrm{abs}$ is limited by the thermal and photorefractive effects \cite{Surya:21}. When the on-chip pump power is relatively high (e.g., over few hundreds of microwatts in this device), the light-induced static space-charge electric field $E_\mathrm{sc}$ will modify both the cavity resonance and quasi-phase matching condition via refractive index change through $\Delta n = -(1/2)n^3r_{33}E_\mathrm{sc}$ \cite{sun2017nonlinear}. Although new quasi-phase matching condition could be satisfied via fine tuning device's temperature, the resonance drifting (occurring at a time scale from 100 $\mu$s to 10 ms) may not be mitigated by self-locking depending on the competition between thermal and photorefrative effect \cite{he2019self}. This calls for an active fast feedback control with bandwidth over 100 kHz to stabilize the highly efficient nonlinear conversion process amid the aforementioned effects. To this end, Pound–Drever–Hall (PDH) locking is one of the popular methods for optical cavity stabilization \cite{black2001introduction}. The primary challenge in implementing PDH stabilization, however, is with the need for high-speed phase modulation, whose bandwidth must be comparable or larger than cavity linewidth (about GHz level). Fortunately, this requirement is well accommodated in thin-film lithium niobate, for its superior nonlinear and electro-optic properties. It allows to integrate the essential optical components, i.e. high-speed phase modulator and efficient frequency conversion microring, on the same chip. In the next section, we will describe in detail the feedback control and resulting performance. 

\newpage

\subsection{Pound–Drever–Hall error signal generation with integrated phase modulator}
\label{PDH}
\begin{figure}
  \includegraphics[width=\linewidth]{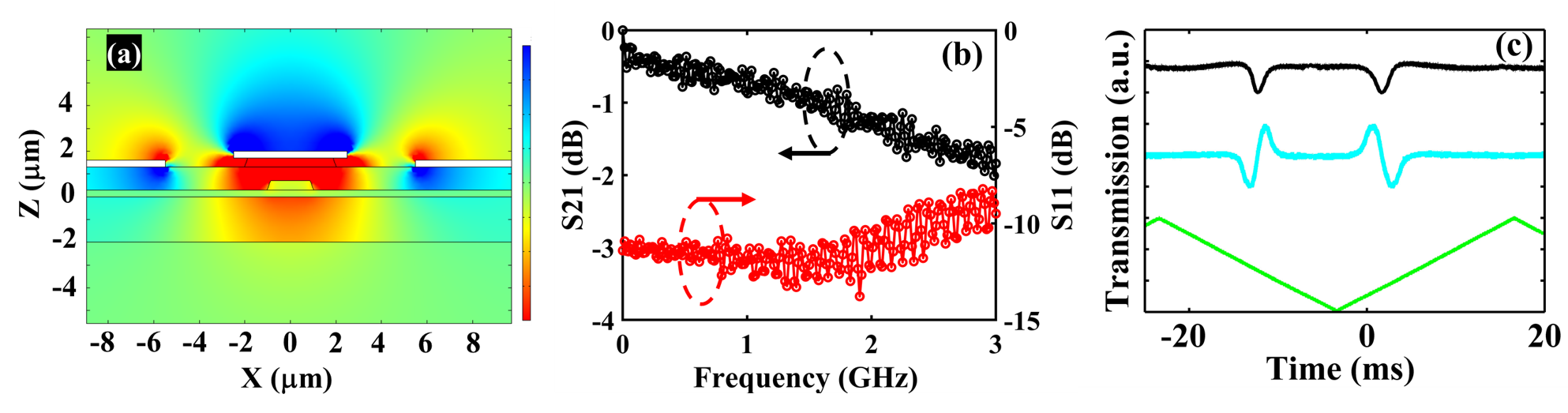}
  \caption{(a) Simulated electrical field across the lithium niobate waveguide. (b) Frequency response of the electrodes, S21: the RF transmission (normalized by subtracting the cable losses) and S11: the RF reflection. (c) The transmission spectra in telecom band (top black line) and the Pound–Drever–Hall error signal (middle canon line) and the applied triangle shape signal (bottom green line). The modulation frequency is about 1 GHz and the on-chip laser power is about -10 dBm. }
  \label{fig3}
\end{figure}

Different from electric-optic modulation using X-cut TFLN \cite{jin2019high}, we devise a new electrode configuration for Z-cut orientation wafer, so as to access its largest electro-optic coefficient $r_{33}$. As shown in Fig.~\ref{fig3} (a), we employ a ground-signal-ground (GSG) scheme rather than the standard ground-signal scheme to enhance our modulation by 1.3 times, where the signal pad is placed on top of the LN waveguide and the two ground pads located on each side. According to the electrical field simulation using COMSOL Multiphysics, the voltage-length product ($V_{\pi}$ $\cdot$cm) is estimated to be about 10 V$\cdot$cm, which is comparable to commercial products (e.g., Thorlab $V_{\pi}$ $\cdot$ cm $>$ 20 V $\cdot$ cm). With this integrated phase modulator, we reduce the system complexity and avoid insertion loss that is typically 4 to 5 dB. Meanwhile, high frequency operation is important for current cavity linewidth (1.25 GHz for $Q_\mathrm{L} \approx 1.5 \times 10^5$). To characterize the electrical properties of the RF electrodes, we measure the S21 and S11 parameters using a 3-GHz vector network analyzer (VNA: Agilent 8714ES). As shown in Fig.~\ref{fig3} (b) and (c), it indicates good impedance matching of the electrodes through 3 GHz, as limited by our VNA bandwidth. Then we use the experiment setup in Fig.~\ref{fig1} (a) to examine the PDH error signal. Firstly, 1 GHz sinusoidal modulation signal is created in the RF signal generator (Agilent 8648C) and then amplified by RF amplifier (26 dBm, Optilab, MD-20-M). It is split into two paths: one serves as local oscillator (LO) and another is applied on the chip via a high-speed GSG probe (GBB, 40A-GSG-150-P). The laser light, after phase modulation, is coupled out and detected by high-speed amplified photodetector (PD2, Sumitomo, ERP1402GT), whose output is mixed with the LO via a mixer (Marki Microwave, M1-0012). Meanwhile, another amplified photodetector (PD1, Thorlab PDA10CS) is used to monitor transmission spectrum. The demodulated signal is sent to a NIST lockbox for locking \cite{leibrandt2015open}. As shown in Fig.~\ref{fig3} (d), s clear PDH error signal is observed when the telecom laser is swept across the resonance around 1561.3 nm. The $\pm1.5$V triangle shape modulation signal is generated via the lockbox. This indicates our on-chip phase modulator is functioning properly and ready for the following locking experiment. 

\subsection{Stabilization of efficient second-harmonic generation}
\begin{figure}
  \includegraphics[width=\linewidth]{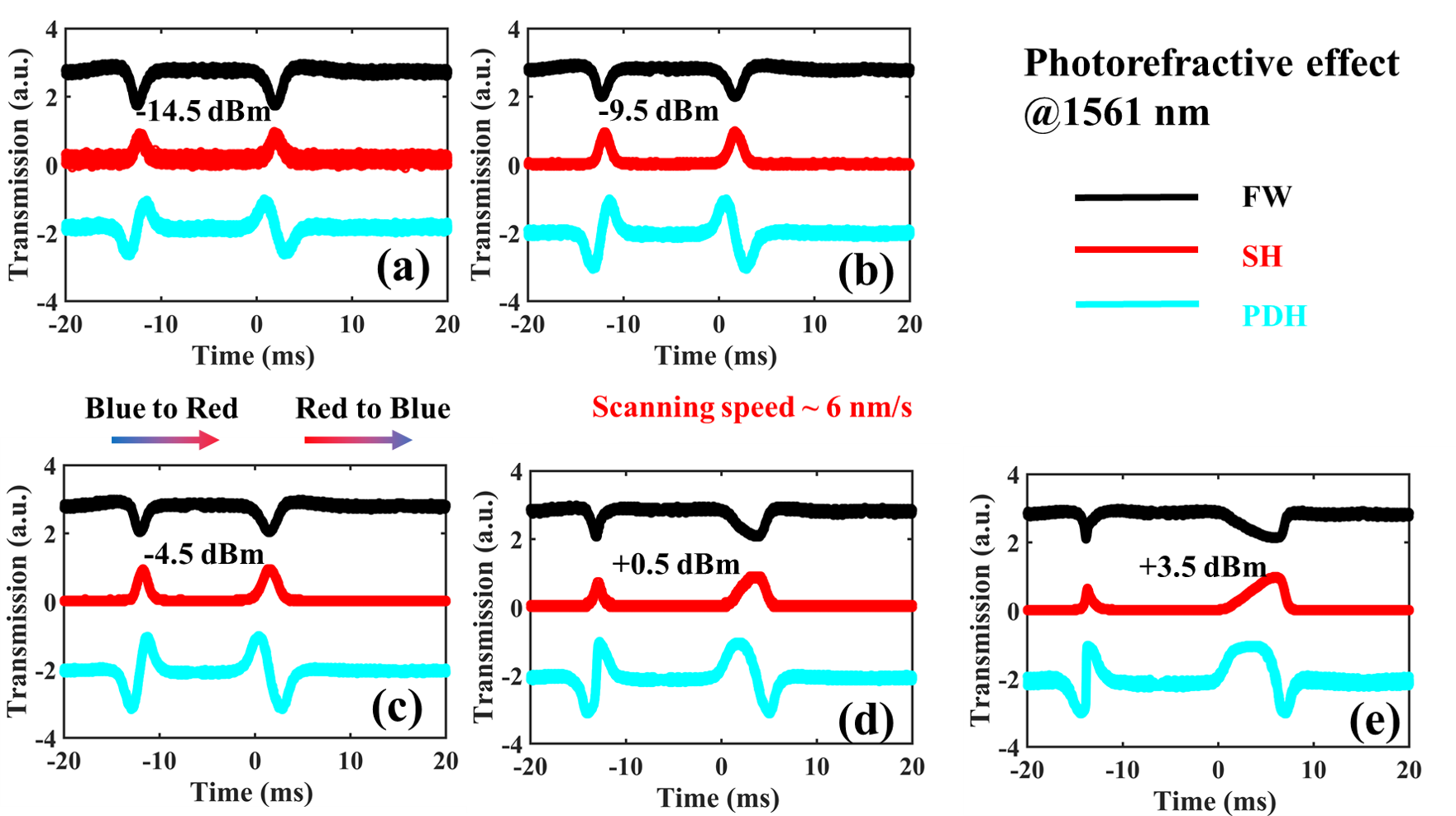}
  \caption{The normalized transmission spectra with various on-chip pump power for telecom cavity mode (top black line), the SH light (middle red line) and the PDH error signal (bottom cyan line) at a fix scanning speed of 6 nm/s. }
  \label{fig4}
\end{figure}

The tight mode confinement and high quality factor of the PPLN microring are the key to achieve 50$\%$ conversion efficiency with only milliwatt pump power. However, these two also enhance the undesirable PR and thermal effects during the nonlinear process. This is because the intracavity light intensity is largely boosted to generate strong electrical field, resulting in detrimental space-charge electric field $E_\mathrm{sc}$. To study the PR effect on SHG, we bidirectionally scan the telecom laser across the phase matching resonance and monitor the telecom and visible transmission spectra, as shown in Fig.~\ref{fig4}. Meanwhile, to prepare for the locking, we also monitor the PDH error signal. At the scanning speed 6 nm/s, we observe the PR effect when on-chip laser power is above -4.5 dBm (see Fig.~\ref{fig4}(d) \& (e)). According to its response under bidirectional scanning, we confirm that the PR effect dominates the thermal effect \cite{wang2019monolithic} with on-chip power up to +3.5 dBm . Note that, here the threshold of the PR effect is over 10 times higher than previous reported results in X-cut LN microring at a similar scanning speed \cite{Xu:21}. This could be explained by different orientations of LN wafers are used, a higher chip temperature (75 $^\circ$C), or the over-coupling condition for cavity modes, but a future study is needed. 

\begin{figure}
  \includegraphics[width=\linewidth]{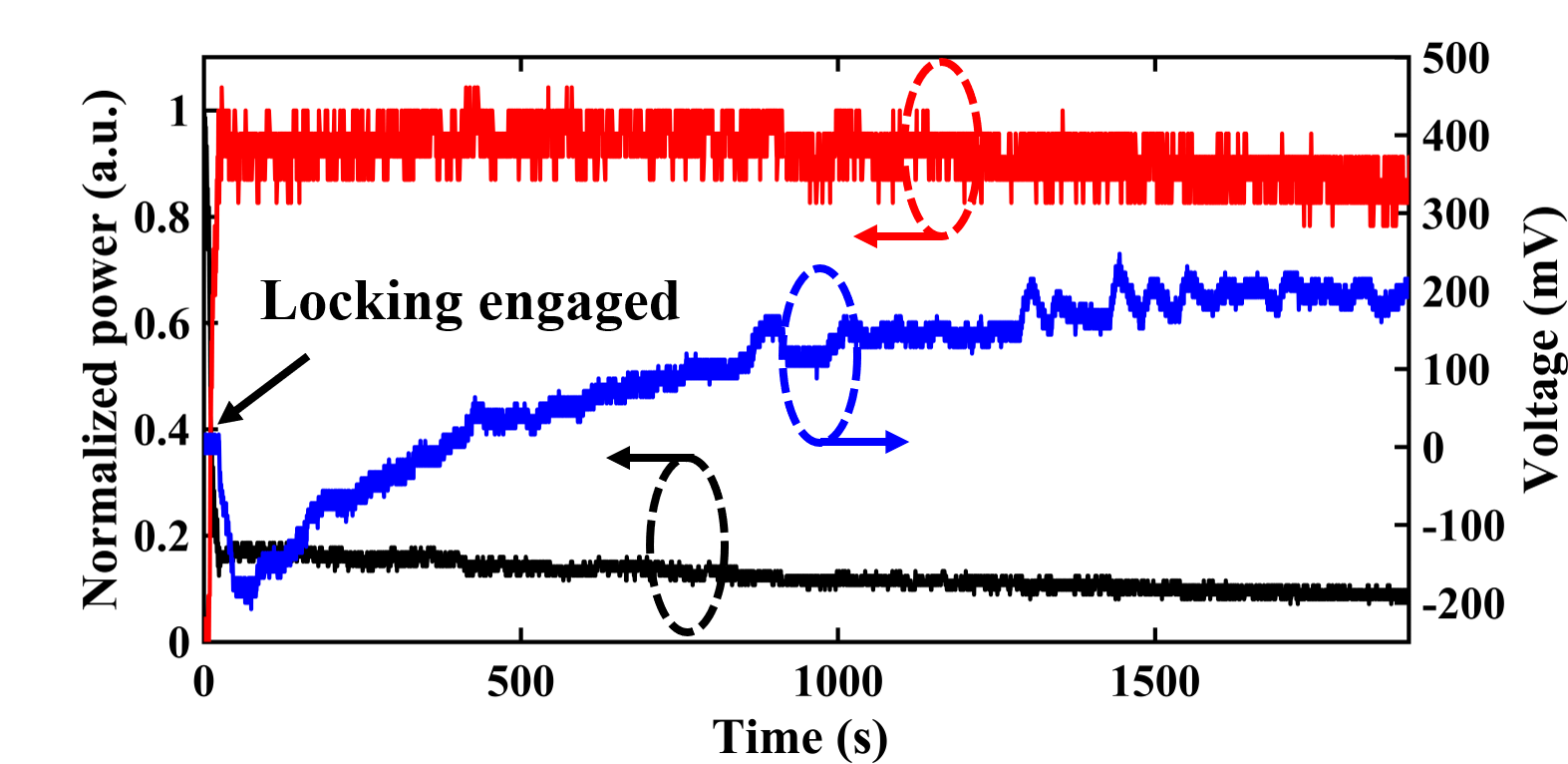}
  \caption{The normalized power of telecom pump (black line), SH light (red line), and the required voltage for the feedback signal (blue line). The locking is engaged after manually tuning the pump close to the cavity resonance. The initial absolute conversion efficiency is 48$\%$ with 1.1 mW on-chip pump power.}
  \label{fig5}
\end{figure}

From the PDH error signal in Fig.~\ref{fig4}, we find that the dynamic range of the PR effect is on the order of millisecond, which agrees with previous studies \cite{sun2017nonlinear}. The induced frequency shift ($\sim$GHz) is within the laser piezo fine-tuning range ($\pm$8 GHz), as indicated in the same figure, even at very high pump power level. Besides, with faster scanning (GHz/$\mu s$) through the amplitude modulation port of Santec laser, the PR-induced frequency shift is suppressed well within laser's tuning range ($\pm$1 GHz). This is essential as the tuning bandwidth need to be a few times larger than PR induced frequency shift. In order to achieve high PDH locking performance, we use the amplitude modulation port of the laser whose speed can reach 400 KHz. To stabilize the pump power within $\pm 0.2 $ dB over the locking process, we flat the modulated laser power by using an EDFA with the fixed output, as shown in Fig.~\ref{fig1}. Before the locking is engaged, we manually approach the resonance from red to blue and start the engaging when close to the resonance. As shown in Fig.~\ref{fig5}, the absolute conversion is stabilized between 48$\%$ and 41$\%$ over 30 minutes. The slight drop over time is mainly due to the coupling instability, because the transmission curve of pump power is drifting downwards accordingly while the feedback signal is kept feeding into the laser. In the future, another feedback loop to the EDFA could be implemented to stabilize the on-chip pump power against the drifting of the coupling. Besides, an additional phase modulation component could be added on top of the microring to achieve frequency tuning \cite{Stefszky:21}, to replace current frequency tuning by the external laser.


\section{Conclusion}
In summary, we have demonstrated highly efficient and actively stablized frequency doubling, through the optimally integrating a Z-cut phase modulator and a PPLN microring on the same chip, with simultaneous overcoupling for both fundamental and second harmonic modes. Enabled by TFLN's ferroelectricity, strong second-order nonlinearity and excellent electro-optic property, such monolithically integrated devices will serve for practical quantum frequency conversion, frequency comb generation and quantum photon sources, etc., where power efficiency, good scalability, and long stability are the key.

\newpage


\medskip
\textbf{Acknowledgements} \par 
Thanks for Mingming Nie's suggestions on the PDH locking implementation. The research was supported in part by National Science Foundation (Award \#1641094 \& \#1842680) and National Aeronautics and Space Administration (Grant Number 80NSSC19K1618). Device fabrication was performed in Nanofabrication Facility at Advanced Science Research Center (ASRC), City University of New York (CUNY).

\medskip
\textbf{Conflict of Interest} \par 
The authors declare no conflict of interest.

\medskip

\end{document}